\documentclass[10pt,preprint]{aastex}

\newcommand{\cc}{\mathrm{cm^{-3}}}

\newcommand{\kmps}{\mathrm{km~s^{-1}}}

\shorttitle{Methanol masers in G9.62+0.20E and G188.95+0.89}
\shortauthors{van der Walt}

\begin{document}


\title{On the methanol masers in G9.62+0.20E and G188.95+0.89}


\author{D.J. van der Walt}
\affil{Centre for Space Research, North-West University, Potchefstroom,
  South Africa}
\email{johan.vanderwalt@nwu.ac.za}

\begin{abstract}
A comparison between the observed light curves of periodic masers in
G9.62+0.20E and G188.95+0.89 and the results of a simple colliding-wind
binary model is made to establish whether the flaring and other
time-dependent behaviour of the masers in these two star forming regions
can be ascribed to changes in the environment of the masers or in the
continuum emission from parts of the background \ion{H}{2} region. It is
found that the light curves of widely different shape and amplitude in
these two objects can be explained within the framework of a periodic
pulse of ionizing radiation that raises the electron density in a volume
of partially ionized gas against which the masers are projected. It is
also shown that the decay of the 11.405 $\kmps$ maser in G188.95+0.89
can be explained very well in terms of the recombination of the ionized
gas against which the maser is projected while it would require very
special conditions to explain it in terms of changes in environment of
the maser. We conclude that for G9.62+0.20E and G188.95+0.89 the
observed changes in the masers are most likely due to changes in the
background free-free emission which is amplified by the masers.

\end{abstract}



\section{Introduction}

Since the discovery of the bright and widespread class II methanol
masers at 6.7 GHz by \citet{menten1991}, research on these masers has
grown enormously. Over the last almost two decades it has been
established rather firmly that the class II methanol masers are
exclusively associated with high mass star forming regions \citep[see
  eg.][]{ellingsen2006}. However, in spite of the large volume of work
that has been done, a clear picture of exactly where in the
circumstellar environment the class II masers arise does not yet exist
\citep[see eg.][]{vanderwalt2007}. This lack of knowledge makes it
difficult to fully exploit the masers to study the circumstellar
environment of the young massive stars with which they are associated.

A significant recent discovery is that a small number of masers show
periodic or regular variability at 6.7, 12.2, and 107 GHz
\citep{goedhart2003, goedhart2004,goedhart2007, goedhart2009,
  vanderwalt2009}. Except for the detection of quasi-periodic flaring of
formaldehyde masers in IRAS 18566+0488 \citep{araya2010} no similar
variability has yet been detected for other masing molecules associated
with high-mass star forming regions. This suggests that for some reason
yet unknown, the class II methanol masers in some star forming regions
are sensitive to a periodic/regular phenomenon associated with these
massive star forming regions. On the one hand, the maser variability
might be due to changes in the physical conditions in the masing region
itself which affect the population inversion. On the other hand the
variability may be due to changes in the background radiation field that
is amplified. Given the regularity of the flaring of these masers, it is
reasonable to assume that only one of the above mechanisms is
responsible for the observed periodic behaviour of these maser. Whether
it is the same mechanism that is responsible for the periodic flaring of
the masers in all the periodic sources is not yet clear.

In a recent paper \citet{araya2010} suggested that not only the flare
events traced by formaldehyde and methanol masers IRAS 18566+0408 but
also the methanol masers in G9.62+0.20E could be caused by periodic
accretion of circumbinary disk material in a very young binary
system. These authors also suggested that the less flare-like and lower
amplitude variablity seen in other periodic methanol maser sources would
be expected from accretion of circumbinary material in binary systems
with lower eccentricity. On the other hand, \citet{vanderwalt2009}
suggested that the methanol maser flares seen in G9.62+0.20E might be
related to changes in the background free-free emission which in turn is
due to changes in the electron density caused by a pulse of ionizing
radiation passing through a volume of partially ionized gas against
which the maser is projected. This suggestion by \citet{vanderwalt2009}
was based on the decay part of the average 12.2 GHz flare profile only
and it is therefore not really clear to what extent the time series can
be explained by a simple colliding-wind binary scenario. Although both
\citet{vanderwalt2009} and \citet{araya2010} suggested an underlying
radiative origin for the flaring, these are two different scenarios with
different implications.

In this paper we use a very simple toy model based on some general
aspects of a colliding-wind binary (CWB) system and apply it to the time
series of G9.62+0.20E and the significantly smaller amplitude periodic
methanol maser source G188.95+0.89 (also known as S252 and AFGL 5180 in
the literature). The use of such a toy model is justified at present
since very little or perhaps nothing is known about the existence or not
of a binary system in each of these star forming regions as well as
about the stellar winds of the exciting stars. It is shown that the time
series for the 1.25 $\kmps$ 12.2 GHz maser in G9.62+0.20E is very well
reproduced over six cycles spanning about 1400 days. It is also shown
that the simple model is able to reproduce the time series of the 10.659
$\kmps$ maser in G188.95+0.89 over five cycles spanning about 1800
days. In addition we show that the decay of the 11.405 $\kmps$ maser in
G188.95+0.89 can be explained very well in terms of the recombination of
a hydrogen plasma.

\section{Theoretical framework} 

At present there is no explicit observational evidence indicating that
G9.62+0.20E and  G188.95+0.89 are colliding-wind binary systems
or even just binary systems. Therefore no observational data exists on
the stellar winds and orbital parameters except for the association with
the periodic methanol masers.  Modelling of both systems based on known
values of physical quantities is therefore not possible. To make
progress we adopt the following working hypotheses.

\begin{itemize}
\item G9.62+0.20E and G188.95+0.89 are CWB systems. In the case of
  G9.62+0.20E the more massive star is at least a B1 type star
  \citep{hofner1996} which is the dominant ionizing source that
  maintains the \ion{H}{2} region. For the present simple model the mass
  of the secondary star is not important except for setting the orbital
  parameters. However, it is implicity assumed that the secondary star
  is a non-ionizing star but which is still massive enough to have a
  significant stellar wind. As a {\it starting point} it is also assumed
  that the post-shock gas cools adiabatically implying that $L_{shock}
  \propto 1/r$ \citep{stevens1992, zhekov1994}, where $L_{shock}$ is the
  luminosity of the post-shock gas and $r$ the distance between the two
  stars.

\item The class II methanol masers amplify the free-free emission from
  the background hypercompact \ion{H}{2} region. In the case of
  G9.62+0.20E it is known from the work of \citet{sanna2009} that all
  the 12.2 GHz maser features are projected against the \ion{H}{2}
  region. No single dish or high resolution radio continuum data could
  be found for G188.95+0.89. The positions of the masers relative to a
  possible background ultra- or hypercompact \ion{H}{2} region is
  therefore not known for this region. Since our underlying assumption
  is that the observed flaring is due to changes in seed photon flux
  (the background free-free emission) the amplification of the masers
  is considered to be time independent.
\end{itemize}

\begin{figure}[h]
   \centering \includegraphics[width=12cm,angle=0,clip]{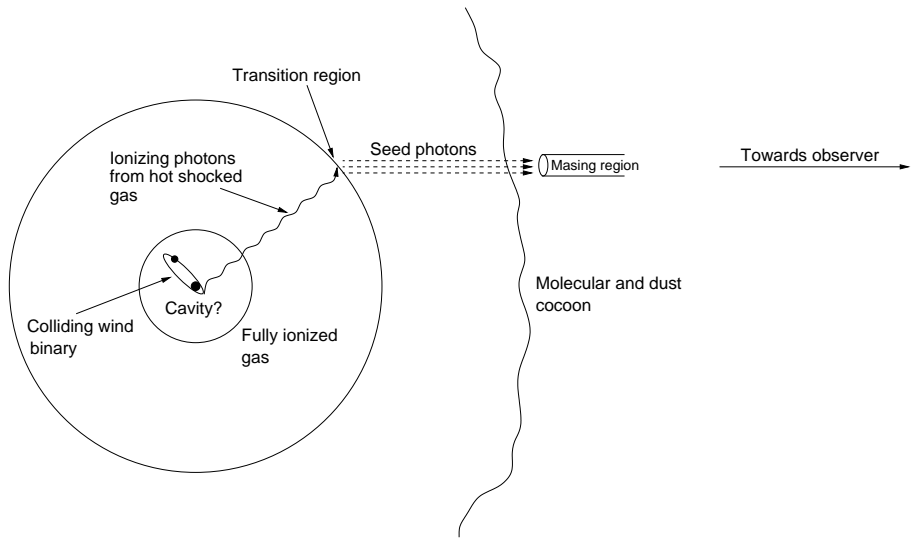}
      \caption{Schematic representation of the CWB scenario to explain
        the periodic masers in G9.62+0.20E and G188.95+0.89. See
        Fig. \ref{fig:ionstruct} for an example of the ionization
        structure of the \ion{H}{2} region.}
\label{fig:cwb}
   \end{figure}

With the above working hypotheses the basic picture is that, whether the
shock heated gas cools adiabatically or radiatively, the orbital motion
results in the modulation of the ionizing radiation originating from the
shock heated gas. This results in a ``pulse'' of ionizing radiation
propagating outwards almost unattenuated through the \ion{H}{2} region
until it reaches substantially partially or non-ionized gas where it is
absorbed. This raises the electron density and consequently the
free-free emission. A schematic representation of the CWB scenario is
given in Fig. \ref{fig:cwb}.

\begin{figure}[h]
   \centering \includegraphics[width=12cm,angle=0,clip]{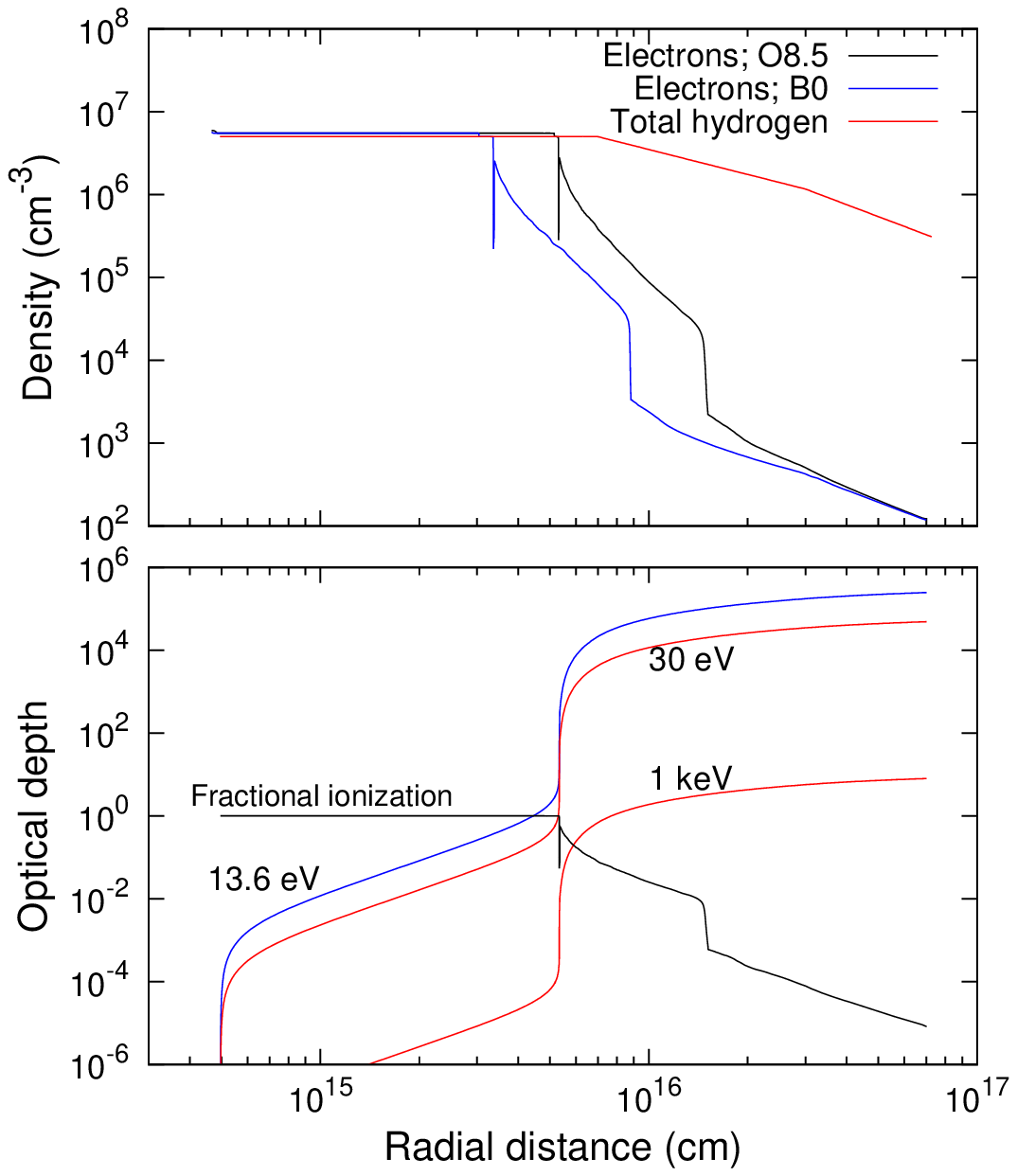}
      \caption{{\it Upper panel:} Examples of the ionization structure
        of a hyper/ultracompact \ion{H}{2} region. {\it Lower
          panel:} Radial dependence of the optical depth for photons of
        different energy for the ionization structure as produced by an
        O8.5 star as shown in the upper panel. }
\label{fig:ionstruct}
   \end{figure}

It is also necessary to expand somewhat on the location of the partially ionized
gas referred to above. Examples of the possible ionization structure of a
hyper/ultra-compact \ion{H}{2} region as calculated with the photo-ionization
code {\tt Cloudy} \citep{ferland1998} are shown in the upper panel of
Fig. \ref{fig:ionstruct}.  The red line gives the radial distribution of
hydrogen while the black and blue lines are for the resulting radial electron
densities for ionizing stars of spectral types B0 and O8.5 respectively. In both
cases it is seen that the fully ionized region ends in a rather rapid decrease
in the electron density. Although the decrease in the electron density is very
rapid, the transition region is still of finite width. Closer inspection shows
that the electron density decreases from a value of $5\times 10^6~\cc$ to $3
\times 10^5~\cc$ over a distance of about $10^{14}$ cm.

Conversely, it means there is a sharp increase in the hydrogen density across
the transition region, implying a sharp increase in the optical depth for
ionizing photons.  The lower panel of Fig. \ref{fig:ionstruct} shows the optical
depth as a function of distance into the \ion{H}{2} region for 13.6, 30, and 1
keV photons using the ionization structure for the case of the O8.5 star in the
upper panel.  It is seen that 13.6 eV photons are absorbed even before reaching
the transition region. The reason for this is that although the fractional
ionization equals one at the inner edge of the \ion{H}{2} region, it slowly
decreases outwards. The neutral hydrogen column density between the inner edge
and the transition region is therefore not zero. For example, for a total
hydrogen density of $5 \times 10^6~\cc$ and a fractional ionization of, say,
0.998, the mean free path of a 13.6 eV photon is only $1.6 \times 10^{13}$
cm. Higher energy, $>$ 30 eV, ionizing photons can, however, reach the partially
ionized gas, while $>$ 1 keV photons can propagate well beyond the transition
region before being absorbed. What is important to note is that while 30 eV
photons are absorbed almost completely in the transition region, 1 keV photons
are absorbed over a much greater distance interval. The transition region
therefore provides a well defined region of partially ionized gas where the
ionization pulse's photons with energy roughly between 30 eV and somewhat less
than 1 keV can produce additional ionization and thereby raise the electron
density. In reality the \ion{H}{2} regions associated with the two periodic
methanol maser sources considered here will have ionization structures different
from the ideal case discussed above and the energy range of photons responsible
for ionization in the transition region may therefore also differ from the above
example. We also note that due to the suggested rather high photon energies
involved, secondary electrons produced in photoionization may also be a source
of ionization associated with the effect of the pulse of ionizing photons.
 
The time-dependence of the electron density at the position in the
partially ionized gas is then given by the solution of 
\begin{equation}
\frac{dn_e}{dt} = -\alpha n_e^2 + (\Gamma_{\star} +
\Gamma_p(t))n_{\mathrm H^{0}}
\label{eq:rate}
\end{equation}
\citep{vanderwalt2009} where $\alpha$ is the recombination coefficient,
$\Gamma_{\star}$ the constant ionization rate at that particular position due to
the diffuse ionizing radiation from the \ion{H}{2} region and $n_{\mathrm H^0}$
the neutral hydrogen number density also at the same position. $\Gamma_p(t)$ is
the time dependent ionization rate at the same position due to the ionization
pulse. Since the production of secondary electrons is directly related to the
ionization pulse, the ionization rate due to secondary electrons can be
considered as included in $\Gamma_p(t)$.  The first term on the right in
Eq. \ref{eq:rate} gives the decrease of the electron density due to
recombinations and the second term the production of electrons due to
photoionizations. Both $n_e$ and $n_{\mathrm H^0}$ are time dependent and are
related through $n_{\mathrm H^0} + n_e = n_{\mathrm H}$ with $n_{\mathrm H}$ the
total hydrogen number density at that position in the \ion{H}{2} region. The
calculation is done using the ``on the spot'' approximation implying $\alpha =
2.59 \times 10^{-13}~\mathrm{cm^3\,s^{-1}}$.  Considering the radial dimension
of the transition region (of the order of a few times $10^{14}$ cm), a standard
calculation shows that it is optically thin at 12.2 GHz (and frequencies above
that) which implies that the emitted flux of free-free emission at 12.2 GHz due
to the ionization pulse's effect is simply proportional to $n_e^2$.

Since we have not performed a full calculation of the interaction of the
stellar winds and of the physical conditions in the shocked gas, it was
necessary to normalize $\Gamma_p$ in some way. This was done by setting
as initial condition (secondary star located at apastron at $t = 0$)
that at the position of the transition region the ionization rate due to
the pulse is given by $\Gamma_p(r_{max}) = \beta\Gamma_\star$, with
$\beta > 0$ and $ \Gamma_{\star} = {\alpha n_{e,\star}^2}/{n_{\mathrm
    H^0}}$. Here $n_{e,\star}$ is the equilibrium electron density as
determined by $\Gamma_\star$ at that position.

The time dependent source of ionizing radiation in the framework of our
working hypothesis is the hot shocked gas created by the colliding
winds. As noted above, it is assumed that the luminosity of the shocks
scales like $r^{-1}$ with $r$ the distance between the two stars. The
time dependence of the ionization rate, $\Gamma_p(t)$, is therefore the
same as that of $r^{-1}(t)$ where $r$ is given by
\begin{equation}
r = \frac{a(1 - \epsilon^2)}{1 + \epsilon \cos \theta}
\end{equation}
and $a$ is the semi-major axis given by 
\begin{equation}
a = \left[\frac{G(m_1 + m_2)T^2}{4\pi^2}\right]^{1/3}
\end{equation}
Here, $G$ is the universal gravitational constant, $m_1$ and $m_2$ the
masses of the two stars, and $T$ the period, which is 244 days in the
case of G9.62+0.20E and 404 days for G188.95+0.89. The time dependence
of $r^{-1}$ was calculated numerically using Kepler's second law in
the form
\begin{equation}
r^2\frac{d\theta}{dt} = \frac{L}{\mu}
\end{equation}
where 
\begin{equation}
\mu = \frac{m_1m_2}{m_1 + m_2}
\end{equation}
and 
\begin{equation}
L = \mu\sqrt{G(m_1+m_2)a(1 - \epsilon^2)}
\end{equation}
is the angular momentum. The time step was taken as one fifth of a
day. The orbital parameters were calculated using stellar masses of
17$\,\mathrm{M_\odot}$ and 8$\,\mathrm{M_\odot}$ following
\citet{vanderwalt2009}.

\section{Comparison with observed periodic maser light curves}

The free parameters involved in the calculation were $n_{\mathrm{H}}$,
$n_{e,\star}$, $\beta$, and the eccentricity, $\epsilon$. To find a fit
to the observed light curves it was first of all necessary to find
combinations of the above free parameter such that the relative
amplitude (= $(S_{max}-S_{min})/S_{min}$) equals that of the observed
light curve. A preliminary exploration of parameter space showed that
there is no unique combination of the free parameters for a given
relative amplitude. In fact, for a given $\epsilon$ there appears to be
a continuum of combinations of $n_{\mathrm{H}}$, $n_{e,\star}$, $\beta$
for a given relative amplitude.  Since very little or even nothing is
really known about the ionization structure and density distributions of
the two \ion{H}{2} regions, the approach followed here was to fix the
total hydrogen density in the volume under consideration and to find the
range of combinations of $n_{e,\star}$, $\beta$, and $\epsilon$ that
gives the correct relative amplitude. Typically the correct value of
$\epsilon$ was quite easy to determine. This was then followed by a
least squares fitting to find the combination of $n_{e,\star}$ and
$\beta$ that best fits the observed light curve after normalizing the
minimum between two flares to the minimum between two flares in the
observed time series.

\begin{figure}[h]
   \centering \includegraphics[width=12cm,angle=0,clip]{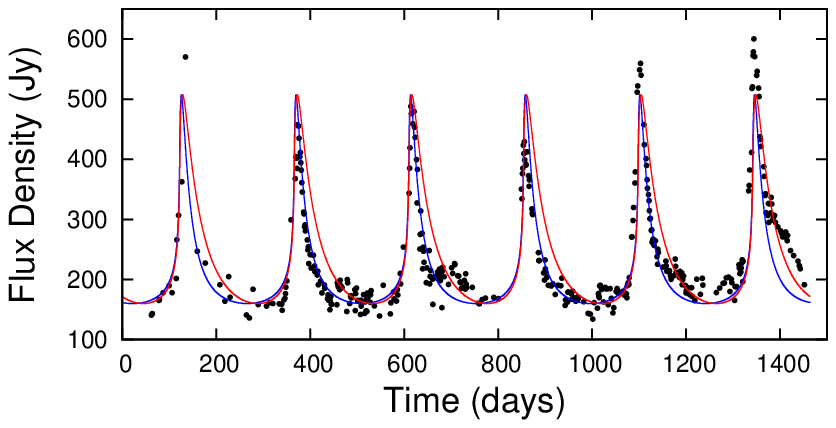}
      \caption{Comparison of the observed 12.2 GHz time series of the
        1.25 $\kmps$ feature and the model time
        series. The relative amplitude is equal to 2.17. Blue: $\beta =
        0.85$, $n_{e,\star} = 7.5 \times 10^5~\cc$. Red:
        $\beta = 2.07$, $n_{e,\star}=3 \times 10^5~\cc$. Day 0
        corresponds to MJD 51000. }
\label{fig:g962}
   \end{figure}

{\it G9.62+0.20E:} The best fit to the time series was obtained with
$\mathrm{n_H}$ fixed at $5 \times 10^6~\cc$, and with $\epsilon = 0.9$,
$\mathrm{n_{e,\star} = 7.5 \times 10^5~cm^{-3}}$, and $\beta = 0.85$.
This best fit solution is compared in Fig. \ref{fig:g962} with the
observed 12.2 GHz time series of the 1.25 $\mathrm{km\,s^{-1}}$
feature. We also show another solution for which $\mathrm{n_{e,\star} =
  3 \times 10^5~cm^{-3}}$, and $\beta = 2.07$ to illustrate the effect
of a lower value of $n_{e,\star}$. The eccentricity of 0.9 implies that
apastron is at 4.4 AU. 

The best fit solution is seen to reproduce the observed time series
remarkably well, especially if it is noted that it does so over six
flares, spaning about 1400 days. There are obviously other effects
present in the data which the very simple model cannot account for. The
result for $\mathrm{n_{e,\star} = 3 \times 10^5~cm^{-3}}$ is seen to
have the same rise time as for $\mathrm{n_{e,\star} = 7.5 \times
  10^5~cm^{-3}}$, but to have a slower decay. This is due to the fact
that the recombination time for a lower density plasma is longer than
for a higher density plasma. We do not show the result for
$\mathrm{n_{e,\star} = 2 \times 10^6~cm^{-3}}$, for example, but merely
state that, on the other hand, in this case the decay is significantly
shorter than for the best fit with $\mathrm{n_{e,\star} = 7.5 \times
  10^5~cm^{-3}}$.

   \begin{figure}[h]
   \centering \includegraphics[width=12cm,angle=0,clip]{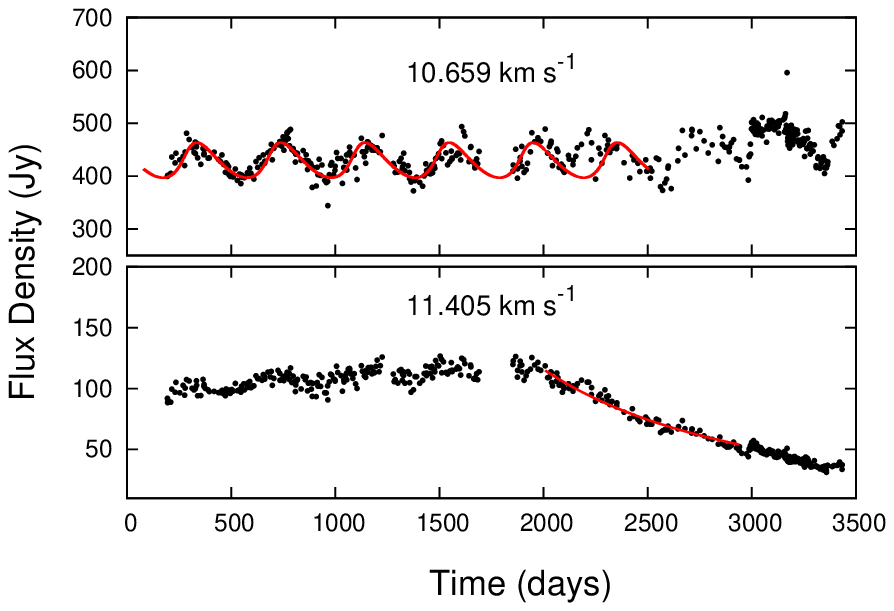}
      \caption{{\it Top panel:} Comparison of the model and observed
        light curves for the 10.659 $\kmps$ periodic maser
        feature in G188.95+0.89. {\it Bottom panel:} Time series of the
        11.405 $\kmps$ maser. The solid red line is a fit of
        eq. A7 of \citet{vanderwalt2009} between days 2007 and 2940. Day
      0 corresponds to MJD 51000.}
\label{fig:g188}
   \end{figure}

{\it G188.95+0.89:} For G188.95+0.89 we used the 10.659 $\kmps$ maser
feature to fit the model result to. The relative amplitude for this
feature is only 0.17 compared to 2.17 for G9.62+0.20E. In this case
$\mathrm{n_H}$ was fixed at $7 \times 10^5~\cc$. The best fit was
found for $\epsilon = 0.36$, $n_{e,\star} = 1 \times 10^5~\cc$, and
$\beta = 0.95$ and is compared in Fig. \ref{fig:g188} with the
observed time series. The model reproduces the observed time series
reasonably well up to about 2000 days after which, for all the maser
features (see \citet{goedhart2007}) there seems to be some general
change in behaviour. In this case the eccentricity of 0.36 and period
of 404 days imply apastron to also be 4.4 AU, similar to that of
G9.62+0.20E. Periastron is about 2.1 AU. The same masses for the
primary and secondary were used as for G9.62+0.20E.

A very interesting aspect of the masers in G188.95+0.89 is that from
about day 2000 the 10.483 $\kmps$ and 11.405 $\kmps$
masers started to become fainter while the other masers stayed more or
less at the same flux level \citep{goedhart2007}. The time series for the
11.405 $\kmps$ feature is shown in the bottom panel of
Fig. \ref{fig:g188}. A small amplitude variation which is in phase with
the 10.659 $\kmps$ maser can been seen for the first almost
1700 days. From about day 2000 there is a decay in the maser flux
density.

The obvious question is what causes such different behaviour for the two
masers in G188.95+0.89 even though most of the maser features are
projected within about 100 AU from each other \citep{minier2002a}.  As
already stated, our basic hypothesis is that the periodic flaring of the
masers might be due to changes in the background free-free emission. We
therefore also tried to fit the decay of the 11.405 $\kmps$ maser with
the expected change in free-free emission from an optically thin
recombining hydrogen plasma as given by Eq. A9 of
\citet{vanderwalt2009}. The data was fitted between days 2007 and 2940
and the result is shown as the solid red line in the bottom panel of
Fig. \ref{fig:g188}. It is seen that for almost 1000 days the decay is
described very well by what is expected of the recombination of an
optically thin hydrogen plasma. In this case the decay is consistent
with the recombination of a plasma with an initial density, $n_{e,0}$,
of about $1.4 \times 10^4~\cc$ recombining to a very low equilibrium
value, $n_{e,\star}$, of less than $10^3~\cc$.

\section{Discussion}

What criticism can be raised against the simple model? It certainly is
true that it contains only some elements of a CWB system and therefore
cannot be regarded as fully realistic. For example, it has been assumed
implicitly that the properties of the winds are such that the shocked
gas has high enough temperature and sufficient volume emissivity to
produce the neccesary ionization at some distant volume in the
\ion{H}{2} region. Furthermore, it has also been assumed that if the
underlying source of the ionization pulse is indeed a CWB, that the
shocked gas cools adiabatically. This need not necessarily be the case
since radiative cooling can result in faster cooling than in the
adiabatic case. 

A second point that might be raised against the present explanation for
the periodic masers is the presence of dust in hyper- and ultra-compact
\ion{H}{2} regions. The shell-like morphology for the mid-infrared
emission for many compact \ion{H}{2} regions as seen, for example, in
the work of \citet{phillips2008} suggests that for compact \ion{H}{2}
regions some of the dust has been cleared out of the ionized volume
either through destruction of the dust or through radiation pressure
(see also \citet{tielens2005}). In the ultra-compact phase, however,
there may still be significant amounts of dust inside the \ion{H}{2}
region. In fact, \citet{debuizer2003} found mid-infrared emission
associated with G9.62+0.20E, suggesting the possible presence of dust
inside the ionized region. Since the presence of dust increases the
optical depth for ionizing photons, it might be argued that the
mechanism proposed here simply might not be realistic since most of the
ionizing photons will be absorbed before reaching the transition region.

The best fit solutions for both G9.62+0.20E and G188.95+0.89
however, suggests that even with the above criticism, the simple model
must contain some element of reality. Even though the assumption that
the shocked gas cools adiabatically has not been justified, it is seen
that the rise of the flares are well described by the $1/r$ dependence
used for the ionization pulse. Although this cannot be considered as
conclusive proof, it is suggestive that the adiabatic assumption might
indeed apply in reality. Overall it is quite remarkable that the
flares are so well described by the solutions of Eq. \ref{eq:rate} that
it is difficult to avoid not to conclude that the flaring of the masers
is indeed due to changes in the free-free emission from those parts of
the background \ion{H}{2} region against which the masers are projected.
 
As for the behaviour of the masers in G188.95+0.89 the question can be
asked how it is possible for some of the masers to still show periodic
behaviour while for one the periodic behaviour seems to have stopped
and its flux density is slowly decreasing, especially when the masers
appear to be located within a projected region as small as 100 AU in
diameter. This question applies not only if it is assumed that the
observed periodic variability and the behaviour of the 11.405 $\kmps$
maser are related to changes in the free-free emission from parts of the
background \ion{H}{2} region but also if assumed that the variability is
due to changes in the maser amplification.

In this regard it has to be noted that if the decrease of the flux density of
the 11.405 $\kmps$ maser is indeed due to changes in the amplification of the
maser, the implication is that the well defined decay over 1000 days imposes
rather strict constraints on the changes in the maser optical depth.  It has
already been shown that the decay is consistent with the recombination of the
ionized gas with an initial electron density ($n_{e,0}$) of about $1.4 \times
10^4~\cc$ to an equilibrium density ($n_{e,\star}$) of less than
$10^3~\cc$. Thus, $n_{e,\star} \ll n_{e,0}$, implying, according to Eq. A8 of
\citet{vanderwalt2009}, that for the maser flux density $S(t) \propto (1 +
\alpha n_{e,0} t )^{-2} \approx \exp(-\alpha n_{e,0} t)$. The observed decay of
the maser is therefore approximately exponential. If this decay is due to a
change in the amplification of the maser then also $S(t) \propto \exp(\tau(t))$,
which together with the observed exponential decay requires that $\tau$ has to
decrease linearly with time for the 11.405 $\kmps$ maser. Since the optical
depth depends on the level populations which in turn depends, amongst other
things, on the pumping radiation field, it is seen that the exponential decay of
the 11.405 $\kmps$ maser implicitly imposes rather strict constraints on how the
pumping radiation field in the maser environment should depend on time such that
$\tau$ decreases linearly with time. It is unclear what physical process or
combination of processes in the star forming environment can be tuned so finely
to result in the optical depth of one of the masers decreasing linearly over
1000 days. On the other hand, since the decay of the maser follows the behaviour
of what can be expected for the free-free emission from a recombining plasma, it
seems reasonable to conclude that it reflects the behaviour of the background
free-free emission.

It can also be argued that while the 11.405 $\kmps$ maser might reflect
the behaviour of the free-free emission of the background source, the
periodic masers are due to periodic changes in the amplification. Again
this would mean that the pumping radiation field should have some very
specific time dependence to produce the observed periodicity.  Also,
since the masers are pumped radiatively by infrared photons at which
wavelengths the cloud is most likely internally optically thin on the
100 AU scale, it is hard to envisage a scenario whereby the 11.405
$\kmps$ maser is to such an extent shielded from the general
time-dependent pumping radiation field to not show any periodic
behaviour while the other masers are still subject to the pumping
radiation field and continues to show periodic behaviour. 

Considering the above it seems that the most probable physical situation to
understand the behaviour of all the masers in G188.95+0.89 in a consistent way
is that they reflect changes in the background free-free emission amplified by
the masers.  This means that the underlying cause for the decay of the 11.405
$\kmps$ maser should be related to some physical event in the \ion{H}{2} region
that most probably reduced the flux of ionizing photons reaching the volume of
gas against which the maser is projected, thereby leading to the recombination
of the ionized gas. Since the masers amplify emission from a relatively small
part of the background \ion{H}{2} region, this could have been a very local
event that did not affect other parts of the \ion{H}{2} region. Without any
other observational evidence it is, however, not possible to make any conclusive
statements as to exactly what kind of event gave rise to the onset of the
recombination of the background ionized gas and thus the decay of the maser. In
this regard we point out that \citet{franco2004} detected a decrease in the
radio flux density from the lobes of the bipolar \ion{H}{2} region NGC 7538 IRS1
and suggested the inflow of gas from a neutral torus into the core region as a
possible explanation.

Evaluating whether the circumbinary accretion model proposed by
\citet{araya2010} can also explain the maser flares and other
time-dependent behaviour of masers in G9.62+0.20E and G188.95+0.89 as
presented above, falls outside the scope of this paper. However, the
following two questions are raised. First, if the background free-free
emission remains constant and the maser variability is driven by a
variable infrared pumping radiation field, it is expected, as already
argued above, that all the masers for a specific object will show the
same type of variability especially if it is taken into account that,
for example in the case of G188.95+0.89, they all lie in a volume with a
diameter of about 100 AU. Clearly this is not the case for G188.95+0.89
where the 11.405 $\kmps$ maser shows a completely different behaviour
compared to the 10.695 $\kmps$ maser.

One of the reasons why \citet{araya2010} proposed the circumbinary
accretion model is that the time dependence of the accretion rate seems
to be similar to that of the maser flare profiles.  The second question
concerns to what extent the maser flare profile, even for a saturated
maser, will indeed resemble the original time dependence of the
accretion rate. In the simple CWB model used here, the ionizing pulse
does not resemble the maser flare profile at all, but is ``transformed''
through its interaction with the partially ionized gas. In view of the
fact that a number of different physical processes link the population
inversion in the masing region to the accretion process, is it not clear
whether there will be a one-to-one correspondence between the final
maser flare profile and the time dependence of the accretion rate.

In this regard we note the following: If it is assumed that within the
circumbinary model the eccentricity for G188.95+0.89 is the same as
within our simple CWB model, the time dependence of the accretion rate
should be somewhere between the two cases presented in Fig. 2 by
\citet{artymowicz1996}. It can therefore be expected that, as in the
case for an eccentricity of 0.1, there will still be a significant time
lag in the accretion rate onto the two stars. Since the heating of the
dust depends on the emission from the accretion shocks associated with
{\it both} stars, it is necessary to consider the time dependence of the
{\it total} accretion rate and not that of the individual stars. Due to
the time lag between the accretion rates of the two individual stars,
the time dependence of the total accretion rate in the system for an
eccentricity of 0.1, will therefore not have the sinusoidal-like pattern
that applies to the two stars individually but will deviate
significantly from it.  Although the time lag decreases with increasing
eccentricity, it is expected that there will also be a significant time
lag for an eccentricity of about 0.35 resulting in the total accretion
rate not having a sinusoidal character. The question that obviously
follows is how, for low eccentricity binaries, the accretion rate and
the maser flare profile are coupled?

\section{Conclusions}

In view of the comparison between the model results and the observed
maser flares we conclude that the simple toy model based on aspects of a
CWB system can explain the different flaring behaviour of the masers in
G9.62+0.20E and G188.95+0.89. It was also shown that the decay of the
11.405 $\kmps$ maser in G188.95+0.89 can be explained very well in terms
of the recombination of ionized gas. For G188.95+0.89 it seems as if the
behaviour of all the masers in this source can be understood in terms of
changes in electron density in different parts of the background
\ion{H}{2} region. In the case of G9.62+0.20E the simple CWB model
reproduce the flares of 12.2 GHz masers for six consecutive
flares. Qualitatively the CWB model can also explain the slow increase
of some of the 6.7 GHz masers in G9.62+0.20E. Also in this case does it
seem that the behaviour of the masers can be understood in terms of
changes in the electron density in different parts of the \ion{H}{2}
region.

The working hypothesis under which the above calculations were made was
that both systems are colliding-wind binaries. The question is
whether it can be concluded from the above comparisons of the model
result with the data that this is indeed the case. Our simple toy model
does not really answer this question. Apart from the orbital motion
which determines the pulse shape and the physics involved in
Eq. \ref{eq:rate}, we have not truely modeled a CWB system. More
detailed modeling involving the interaction of the winds, the
temperature distribution of the shocked gas, and the calculation and
propagation of the spectral energy distribution of radiation emitted by
the hot shocked gas are aspects which still need to be done. The fact
that it seems possible to explain widely different masers light curves
such as that of G9.62+0.20E and G188.95+0.89, is sufficient
motivation to persue such more detailed calculations. 

\acknowledgments

The data on G9.62+0.20E and G188.95+0.89 presented here were originally
collected by Mike Gaylard and Sharmila Goedhart. This work was supported
by the National Research Foundation under Grant number 2053475. The author would
like to thank an anonymous referee for constructive comments.

\bibliographystyle{apj}
\bibliography{ref}

\end{document}